\documentclass[aps,pra,showpacs,twocolumn]{revtex4}
\usepackage[english]{babel}
\usepackage[T1]{fontenc}
\usepackage{amsmath,amsfonts,amssymb,float,graphicx,epsfig,color,times,bbm,bm}

\begin{document}
\title{Motional Entanglement with Trapped Ions and a Nanomechanical Resonator}
\author{F. Nicacio}
\affiliation{Instituto de F\'\i sica ``Gleb Wataghin'', Universidade Estadual de Campinas,  13083-970, Campinas, S\~ao Paulo, Brazil}
\author{K. Furuya}
\affiliation{Instituto de F\'\i sica ``Gleb Wataghin'', Universidade Estadual de Campinas,  13083-970, Campinas, S\~ao Paulo, Brazil}
\author{F. L. Semi\~ao}
\affiliation{Centro de Ci\^encias Naturais e Humanas, Universidade Federal do ABC, 09210-170, Santo Andr\'e, S\~ao Paulo, Brazil}
\begin{abstract}
We study the entangling power of a nanoelectromechanical system (NEMS) simultaneously interacting 
with two separately trapped ions. 
To highlight this entangling capability, we consider a special regime where the ion-ion coupling does not generate 
entanglement 
in the system, and any resulting entanglement will be the result of the NEMS acting as an entangling device. 
We study the dynamical behavior of the bipartite NEMS-induced ion-ion entanglement as well as the tripartite 
entanglement of the whole system (ions+NEMS). 
We found some quite remarkable phenomena in this hybrid system. 
For instance, the two trapped ions initially uncorrelated and prepared in coherent
states can become entangled by interacting with a nanoelectromechanical
resonator (also prepared in a coherent state) as soon as the ion-NEMS coupling achieve a certain value, and
this can be controlled by external voltage gate on the NEMS device. 
%
%
We also show that  dynamically the tripartite entanglement presents a more pronounced robustness against the 
destructive effects of dissipation when compared to the bipartite content.
\end{abstract}
\maketitle
\section{Introduction} 
During the last years, theoretical and instrumental developments in physics have led to an unprecedented level of 
control over atomic and nanoscale systems. Striking examples include the manipulation of single quantum systems 
such as laser cooled trapped ions \cite{ions_review}, and the advances on fabrication, characterization and 
application of nano- mesoscale devices such as electromechanical resonators \cite{nems_review}. 
Isolated single quantum systems are the natural choice when quantum logic is involved \cite{implem_review} and the 
nanoscale devices are recognizedly known as very sensitive weak signal detectors \cite{ss,ssi,detec}. 
The emerging field of quantum technology takes advantage of the strong and different features of these 
two kind of systems by combining them together in a single setup \cite{tz,m,s_n}. 

The goal of quantum technology is to explore legitimate quantum resources such as entanglement 
to perform tasks with no classical analogue and to enforce it on nano- mesoscopic systems which 
can be integrated, for instance, on a chip for massive use. In order to achieve that, one very 
attractive route consists of cooling the NEMS down to its ground state \cite{teufel} and then enforce on it a 
legitimate quantum behavior by coupling it to well controlled small quantum systems. 
For example, the coupling of qubits to these mechanical nanoresonators allows one to generate 
superpositions of macroscopically distinguishable states in the motion \cite{cat}. 
These states may be useful to study the so called quantum-to-classical transition and the phenomenon of 
decoherence happening in nano- and mesosystems. This is a natural follow up to the studies started in 
the scope of cavity quantum electrodynamics \cite{har}. 
On the other hand, potential applications of NEMSs include several uses as a sensitive detector. 
Detection of single spins \cite{ss} and spin-spin interactions \cite{ssi} as well as single molecule mass 
spectrometry \cite{smms} offer new possibilities in chemical imaging and characterization. 
For instance, nitrogen-vacancy centers in a diamond tip coupled to NEMS resonators operating at 
room temperature \cite{nv} offers a new route to detect a chemical element that carry a nuclear magnetic moment. 
This device is sensitive enough to determine the element identity and arrangement in a complex molecule. 
In the case of mass spectroscopy, NEMS resonators are believed to enable in the future a detection device 
operating with resolution better than 1 Dalton (the hydrogen atom mass) \cite{nrev}. 
Nowadays, the state-of-the-art is that single gold atoms can be measured \cite{nrev}. 

Trapped ions are well studied quantum systems where experimental control of coherent quantum 
dynamics, state preparation, and measurement are achieved with great precision and good fidelity \cite{ions_review}. 
As a matter of fact, trapped ions have been demonstrated to achieve an error probability per randomized single-qubit 
gate below the threshold estimate commonly considered sufficient for fault-tolerant quantum computing \cite{ts}. 
This achievement of low single-qubit-gate errors is an essential step toward scalable quantum computers and 
simulators. 
It is then clear that a hybrid system composed of trapped ions and NEMS resonators represent an interesting 
platform to study the interplay of quantum mechanical effects in atomic and nanoscopic systems. 
Following this idea, one proposal \cite{tz} suggest the use of a trapped ion as a probe or device controller 
acting on the NEMS. In another proposal \cite{m}, trapped ions are applied to monitor and manipulate the number 
state of a NEMS resonator such that it enables a statistical inference
of the mean phonon number of the oscillator. 

Here, we take a step forward and propose the use of a NEMS resonator 
to generate entanglement between two trapped ions.
This is a conceptually important step since it goes the other way around of what is normally performed. 
Instead of using the microsystem to cause quantum behavior of the mesoscopic system, we use the latter to 
enforce quantum correlations on the former. Since a classical resonator is unable to generate entanglement 
among quantum systems 
coupled to it, this kind of study may work as a kind of classicality test in what concerns entanglement 
generation. 
We also study the qualitative behavior of tripartite entanglement as given by residual tripartite 
entanglement \cite{re} and contrast its dynamical behavior to the ones of the bipartite entanglements 
\cite{nemes}. 
Moreover, we investigate the regime where NEMS is either in an initially thermal state or subjected to energy 
relaxation, where in the former case we see that the tripartite entanglement is 
more robust against energy relaxation than the bipartite entanglement when dynamics is concerned.

The paper is organized as follows. Section \ref{Model} is dedicated to the presentation of the physical system 
studied here and Section \ref{Tools} presents the theoretical tools necessary to understand the results 
presented 
in Section \ref{Results}. Finally, Section \ref{Conclusion} contains a summary of our results and conclusions. 
We dedicate two appendix to present details of some lengthy analytical expressions and further numerical 
analysis.
\section{Physical System and Model}\label{Model}
\begin{figure}[h]
 \centering\includegraphics[width=0.8\columnwidth]{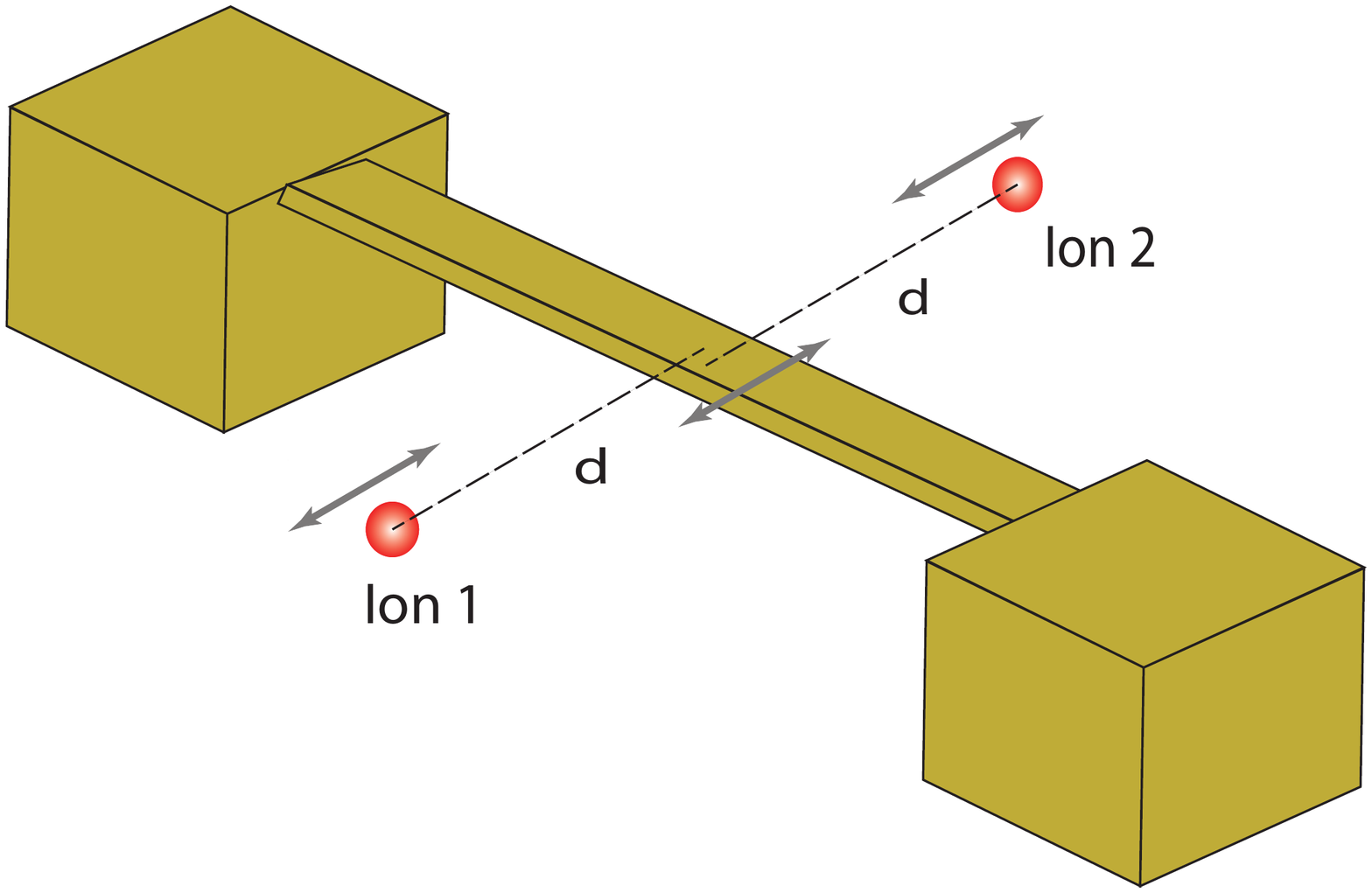}
 \caption{Sketch of the physical setup considered in this paper. 
          Two trapped ions (traps not shown) interact with the flexural motion of a doubly 
          clamped nano beam which is charged by a bias gate not shown in the picture.}
 \label{fig1}
\end{figure}

In this work, we consider a system consisting of two separate trapped ions 
capacitively coupled to a small doubly clamped nano beam (a NEMS resonator). 
The system is depicted in figure \ref{fig1}. 
The ions-NEMS coupling results from the application of an external bias voltage at an electrode on the 
NEMS 
(the bias gate is not shown in the picture).

The interaction energy between the motional degree of freedom of an ion with charge $+e$ and a 
NEMS with bias gate charge $Q$ is given by the electrostatic coupling \cite{tz,m}
\begin{eqnarray}
V=\frac{keQ}{|d \pm (\hat{x} - \hat{X}_i)|}
\end{eqnarray}
where $\hat{X}_i$ and $\hat{x}$ refer to position operators of one ion and the NEMS, 
respectively, around their equilibrium positions. At equilibrium, each ion is separated from 
the center of mass position of the NEMS by a distance $d$. Since we are dealing with positions 
around the equilibrium, the signs $\pm$ are necessary to distinguish one ion from the other, 
for example, one can associate $+$ with ion 1 and $-$ with ion 2. 
Typically, 
one finds that $\langle\hat{X}_i\rangle,\langle\hat{x}\rangle<<d$ so that the coupling 
energy can be expanded as 
\begin{eqnarray}\label{Vint}
V \approx  \frac{k e V_0 C_0}{d} \left\{1 \pm \frac{[\hat{X}_i-\hat{x}]}{d}+\frac{[\hat{X}_i-\hat{x}]^2}{d^2}\right\}, 
\end{eqnarray}
with $Q =C_0V_0$, where $C_0$ and $V_0$ are the capacitance and gate voltage, respectively. 
According to (\ref{Vint}), the coupling between the $i_{th}$ ion and the NEMS is just $V= - \chi \hat{X}_i
\hat{x}$,
where $\chi= \tfrac{2 k e C_0 V_0}{d^3}$. The linear and quadratic terms involving $X_i$ or $x$ 
alone have been absorbed into redefinitions of equilibrium positions and oscillators' frequencies.

Since the ions are not neutral particles, they will also couple by means of the same mechanism, 
{\it i.e.}, an interaction term proportional to $\hat{X}_1\hat{X}_2$ \cite{brown11}. 
However, once the equilibrium distance between them is $2d$ (twice the distance to the NEMS) 
and we will be considering identical traps (identical oscillation frequencies), only resonant terms 
in $\hat{X}_1\hat{X}_2$ will be relevant in this weak coupling regime (rotating wave approximation RWA) 
\cite{brown11}.  
Considering $V= - \chi \hat{X}_i\hat{x}$ as the interaction between each ion and the NEMS and a weak RWA
interaction between the ions, the Hamiltonian operator for this tripartite mechanical oscillating system reads $(\hbar=1)$
\begin{eqnarray}\label{hamil}
\hat H &=& \omega \hat{a}_0^\dag \hat{a}_0 +  \sum_{j=1,2} \nu_j \hat{a}_j^\dag \hat{a}_j 
        -  \Omega (\hat{a}^\dag_2 \hat{a}_1+\hat{a}^\dag_1 \hat{a}_2)\nonumber \\ 
       && - \tfrac{\sqrt{2}}{4} ( \hat{a}_0 + \hat{a}_0^\dag ) \sum_{j=1,2} \kappa_j (\hat{a}_j+\hat{a}_j^\dag),
\end{eqnarray}%
where $a_0 (a_0^\dag)$ and $a_i (a_i^\dag)$ are the annihilation(creation) operators for the NEMS and ion 
$i$, respectively. Also,  $\nu_i$ is the vibrational frequency of each ion (trap frequency), 
$\Omega$ is the RWA interaction strength between the ions, $m_i$ and $M$ 
the masses of the $i_{th}$ ion and NEMS, respectively, and 
\begin{eqnarray} \label{kappa}
\kappa_i=\sqrt{\frac{1}{m_i M \nu_i\omega}}\chi
\end{eqnarray}
is the coupling constant for interaction between the NEMS and the $i_{th}$ ion.
We should remark that Hamiltonian (\ref{hamil}), whose dynamics we will study, might appear in other 
interacting bosonic systems. A potential candidate would be the experimentally accessed optomechanical 
system in \cite{thompson} by expanding it to include two cavity modes (corresponding to the ions) interacting 
with a micromechanical oscillatory membrane. Other interesting coupled bosonic systems are discussed in 
\cite{b1}.

It is worthwhile to notice that the vibrational motion of the ions are \textit{directly} 
coupled by means of a beam-splitter-like interaction which is well known to not dynamically 
generate entanglement for direct products of thermal or coherent states \cite{xiang-bin}. 
In this situation, by putting such ions in interaction with the NEMS, initial creation of entanglement will be the 
result of the action of this third party acting as an entangling device. The key element for the initial 
creation of ion-ion entanglement is the presence of terms of the kind 
$\hat{a}_0\hat{a}_i$ or $\hat{a}^\dag_0\hat{a}^\dag_i$ with strength $\kappa_i$ 
providing an \textit{indirect} squeezing interaction between the ions. 
This is an interesting problem because, being an indirect interaction, not all values of $\kappa_i$ 
will lead to squeezing and entanglement in the ionic subsystem. Studying the conditions for this to 
happen is one of the goals of this article. 
As a final remark, it is important to emphasize that Hamiltonian (\ref{hamil}) is quadratic 
in the creation and annihilation operators and then preserves Gaussianity. 
\section{Theoretical Tools}\label{Tools}
In this section, we present the main tools used in this work. Basically, we are going to discuss some 
techniques to treat continuous variable systems in Gaussian states and also measures of entanglement 
suitable to study tripartite systems. 
\subsection{Gaussian Systems}
In order to study the entanglement dynamics that results from the quadratic Hamiltonian (\ref{hamil}),
we represent it in terms of coefficients for the 
momenta and coordinate operators grouped together in the vector 
$\hat R = (\hat x_0, \hat p_0 ,\hat x_1,\hat p_1,\hat x_2, \hat p_2)^\top$  
which contains momentum and position operators ordered in a convenient manner. 
The result is 
\begin{equation}    \label{hamil2}
\hat H = \frac{1}{2} \hat R^\top \mathcal{H} \hat R + K =  
\frac{1}{2}\sum_{i,j=0}^2 (\hat x_i {\bf U}_{ij}\hat x_j + 
                          \hat p_i {\bf T}_{ij}\hat p_j) + K , 
\end{equation}%
%
%
with
\begin{equation}
{\bf U} = 
\left( \begin{array}{ccc}
\omega & -\kappa_1/\sqrt{2} & -\kappa_2/\sqrt{2} \\
-\kappa_1/\sqrt{2} & \nu_1 & -\Omega  \\
- \kappa_2/\sqrt{2} & -\Omega & \nu_2
\end{array} \right), 
\end{equation}
\begin{equation}
{\bf T} = \left(
\begin{array}{ccc}
\omega & 0 & 0 \\
0 & \nu_1& -\Omega  \\
0 & -\Omega & \nu_2
\end{array}\right),\label{mhamil2}
\end{equation}
and, finally, $ K\equiv\tfrac{1}{2}(\omega + \nu_1 + \nu_2)$ is a 
constant. 
\ The operators composing the vector $\hat R$ fulfill the usual canonical 
commutation relation: 
$[\hat R_j,\hat R_k]=i \mathsf{ J }_{jk}$, 
where the symplectic $(6 \times 6)$ diagonal-block matrix $\mathsf{ J }$ is 
given by:
\begin{eqnarray}   \label{jsimp}
\mathsf{ J } =  \left[ \begin {array}{cc}
                 0 & 1 \\
                -1 & 0 \end{array}  \right] \oplus 
                \left[ \begin {array}{cc}
                 0 & 1 \\
                -1 & 0 \end{array}  \right] \oplus 
                \left[ \begin {array}{cc}
                 0 & 1 \\
                -1 & 0 \end{array}  \right]. 
\end{eqnarray}%
In order to explore the entanglement content of the present model, 
we present now (and in next subsection) the necessary mathematical formalism of 
Gaussian CV entanglement theory.

Choosing initial Gaussian states for the three oscillators, 
Hamiltonian (\ref{hamil2}) will preserve Gaussianity and we can extract all 
information about the system from the knowledge of the accompanying covariance matrix 
(CM) ${\bm \gamma}$ whose elements are given by 
\begin{eqnarray} \label{defcov}
{\bf {\bm \gamma}}_{jk}
= \tfrac{1}{2} \langle \hat R_j \hat R_k + \hat R_k \hat R_j \rangle  - 
               \langle \hat R_j\rangle \langle \hat R_k\rangle, 
\end{eqnarray}
where $\langle\,\cdot\,\rangle$ denotes an expectation value. For a bipartite system 
composed of subsystems $A$ and $B$ \footnote{In our case, partition $A$ and $B$ could be, for example, 
                                             the NEMS and both ions, respectively. 
                                             Also, we can trace out the NEMS and consider $A$ 
                                             to be ion 1 and $B$ to be ion 2.}, 
an entanglement measure called logarithmic negativity can be written as a function of 
the covariance matrix as \cite{logneg}%
\begin{equation}
N_{A|B}= - \tfrac{1}{2}\sum_j \ln \left [\min(1, \mu_j^{\!\top_{\!\!B} } )\right],\label{logsym}
\end{equation}
where $\mu_j^{\!\top_{\!\!B} }$ are the symplectic eigenvalues %
\footnote{The symplectic eigenvalues of a positive definite matrix ${\bf V}$ are given by 
the moduli of the (cartesian) eigenvalues of the matrix $i \mathsf J {\bf V} $ 
with $\mathsf J $ in (\ref{jsimp}). A demonstration could be found in \cite{gosson}.} %
of the matrix $ 2 {\bm \gamma}^{\!\top_{\!\!B} }$, 
evaluated after partial transposition of system $B$. %
This partial transposition is achieved by a local time inversion 
in the oscillators pertaining to subsystem $B$: $p\rightarrow - p$. 
For example, if the system $AB$ is composed by \textit{two} oscillators, 
${\bm \gamma}^{\!\top_{\!\!B} }=\mathcal{P}{\bm \gamma} \mathcal{P}$, with 
\begin{eqnarray}\label{parttransp}
\mathcal{P} = {\rm Diag}(1,1,1,-1).
\end{eqnarray}
In order to study the time evolution of entanglement in this system, 
we make explicit use of the fact that \eqref{hamil} preserves the Gaussian character of 
the global state, so that  we just have to calculate ${\bm \gamma}(t)$.
For a system Hamiltonian of the form (\ref{hamil2}), the CM evolves as 
\cite{plenio04b,joao09} 
\begin{equation}\label{ev}
{\bm \gamma} (t) = \mathsf{E}_t {\bm \gamma}_0 \mathsf{E}_t^\top,
\end{equation}
where
\begin{equation}\label{E}
\mathsf{E}_t = \exp\left[\mathsf{ J } \mathcal{H} t\right]                     
\end{equation}
is a symplectic matrix obeying $\mathsf{E}_t^\top \mathsf J \mathsf{E}_t = \mathsf J$. %
As a result, all closed system dynamics is dictated by the eigenvalues of $\mathsf{ J } \mathcal{H}$.
\subsection{Tripartite Entanglement}\label{lte}
Our main goal in this work is to study the dynamics of entanglement
in the present system, especially because it is generated by the coupling of a nanoscale system to 
two atomic ions, forming an interesting platform for quantum technology experiments \cite{tz,m}. 
Bipartite entanglement can be quantitatively addressed in a Gaussian system by using the tools 
developed 
in the last section. We now want to study the build up of genuine tripartite 
entanglement in this hybrid system. If one wants to address this issue quantitatively, 
the way is to start from the well known monogamy relation for pure states \cite{re} and define a 
legitimate tripartite measurement called residual tripartite entanglement by using concurrence or 
logarithmic negativity. In the case of mixed states, this quantification has to be done by means of a 
convex-roof extension \cite{cre}.

For the purposes of our investigation, a semiquantitative approach is sufficient since 
we are not aiming at a particular quantum information protocol where a given value of an 
entanglement measure has a precise meaning. Instead, we are interested in knowing
whether or not there is legitimate tripartite entanglement in our system when some bipartite 
entanglement has vanished and also to investigate the robustness of bi- or tripartite entanglement 
against dissipation in the NEMS. We can then 
follow the same procedure taken in \cite{li11} which is based on multimode inseparability
classification proposed by Giedke et al. \cite{G}. 
In our case this would correspond to check whether or not 
the negativity (\ref{logsym}) vanishes for all bipartitions, {\it i.e.},
by assigning $\{0\}$ to the NEMS and $\{1,2\}$ to the ions, 
we want to check when $N_{i|jk}\neq 0$ for all $i,j,k=0,1,2$ with $i\neq j\neq k$
--- according to \cite{G}, the system is genuinely tripartite entangled only if this is the case. 
Consequently, one can, for instance, study the resilience of tripartite entanglement against dissipation 
by 
investigating the quantity
\begin{eqnarray} \label{tripent}
\tau_{012} = {\rm min} [N_{0|12},N_{1|02},N_{2|01}].
\end{eqnarray}
From the multimode inseparability classification \cite{G} 
it is then clear that genuine tripartite entanglement ceases to exist as soon as $\tau_{012}=0$. 

Each of the negativities, $N_{i|jk}$, in (\ref{tripent}) is calculated similarly to eq.(\ref{logsym}),
where the symplectic eigenvalue is extracted from the partial transposition of the 
$6 \times 6$ CM with respect to the whole subsystem $(jk)$, {\it i.e.}, 
doing a local time inversion in $p_j$ and $p_k$. For example, the partition $1|02$
will have a partially transposed CM as 
${\bm \gamma}^{\!\top_{\!\!B} }=\mathcal{P}{\bm \gamma} \mathcal{P}$, with 
$
\mathcal{P} = {\rm Diag}(1,-1,1,1,1,-1).
$
%
\subsection{Interaction with Environment} \label{IntEnv}

The most general linear Markovian evolution of a density operator 
$\hat \rho$ preserving its positivity and the Gaussianity is described by the master equation
(written in the Lindblad form): 
\begin{equation}
\frac{\partial \hat \rho }{\partial t}  = 
 \frac{1}{ i }
            \left[\hat H, \hat \rho \right] 
-\frac{1}{ 2  } 
      \sum_{k} \left(
        \hat L_k^\dagger \hat L_k         \hat \rho +
        \hat\rho         \hat L_k^\dagger \hat L_k -
      2 \hat L_k         \hat\rho         \hat L_k^\dagger 
                      \right) ,         
\label{lindblad}
\end{equation}
with a \emph{quadratic} Hamiltonian such as (\ref{hamil2}) and linear Lindblad
superoperators $\hat L_k$
\begin{eqnarray}                                                                          \label{linearcond}
\hat L_k = \lambda_k \cdot \mathsf J \hat R ,                   
\end{eqnarray}
where $\mathsf J $ is given by (\ref{jsimp}) and $\lambda_k$ 
is a complex vector.
By defining the superoperators
$                                                                      
\mathsf D  = \rm{Re} \, \Upsilon , \,\,\,  
$ and 
$\Gamma  =   
   \mathsf J \left( \mathcal{H} - \rm{Im} \, \Upsilon \right)$ where
$\Upsilon = \sum_{k=1}^M \lambda_k \lambda_k^\dagger,                              
$
one can show that the covariance matrix (\ref{defcov}) 
obeys now the following equation of motion \cite{wallquist10,nicacio}:
\begin{equation}
\frac{d {\bm \gamma}}{dt} = 
      \Gamma {\bm \gamma}  + 
      {\bm \gamma} \Gamma^\top + \, \mathsf D,
\label{covevII}
\end{equation}
whose solution satisfying the initial condition 
${\bm \gamma}(t=0)= {\bm \gamma}_0$ is given by \cite{gardiner09,wallquist10,nicacio}
\begin{equation}   \label{covev}
{\bm \gamma}(t) = e^{\Gamma t} \,
                  {\bm \gamma}_0  \,
               e^{\Gamma^\top t} +
 \int_0^t dt^\prime 
               e^{\Gamma (t-t^\prime)} \,
                  \mathsf D  \,
               e^{\Gamma^\top (t-t^\prime)}  \, .
\end{equation}
When thinking in terms of Wigner functions in phase space,
all environment influence is represented by $\Upsilon$, with its 
imaginary part (contained in $\Gamma$) responsible for dissipation (the
$\mathsf{J}\mathcal{H}$ term alone is the symplectic evolution) while
${\rm Re} \Upsilon$ is responsible for diffusion.

In our system, the nanoscale motion of the NEMS is certainly more susceptible to 
loss and decoherence than the motion of the trapped ions. 
This happens because the former is subjected to coupling to phonons in the substrate 
where the nanobeam is fixed, while the ions are kept trapped by means of electromagnetic 
potentials in low pressure environments (almost no residual gases) where collisions are minimized. 
Then, we will consider
%
\begin{equation}   \label{lindblad2}                                                       
\hat L_1 = \sqrt{\zeta(\bar N + 1) } \, \hat a_0 
\,\,\, \textrm{ and } \,\,\, 
\hat L_2 = \sqrt{\zeta \bar N } \, \hat a_0^\dagger,
\end{equation}
as well as 
\begin{eqnarray}
\lambda_1 &=& \sqrt{\tfrac{\zeta}{2}(\bar N + 1) }\,(i,-1,0,0,0,0)^\top 
\,\, \textrm{ and } \nonumber \\
\lambda_2 &=& -\sqrt{\tfrac{\zeta}{2} \bar N }\,(i,1,0,0,0,0)^\top,  
\end{eqnarray}%
where $\bar{N}$ is the mean thermal occupation number of the reservoir which is related to the temperature 
of the substract, and $\zeta$ is the damping constant. Even in our case where the Hamiltonian and 
Liouvilian are quadratic on quadrature operators, the 
solution of the problem is quite involved due to matrix exponentiation and integration in (\ref{covev}). 
Next subsection will be devoted to the discussion of some particular regimes where the solution is feasible 
either analytically or numerically.
\section{Results}\label{Results}
In what follows we present our results concerning entanglement dynamics in the hybrid 
ion-NEMS-ion system presented in
last section. We will be considering physically motivated cases consisting of closed and 
open system dynamics with zero or finite temperature $T$. In particular, 
we will be considering identical traps and ion species leading to %
equal frequencies for the ions $\nu_1 = \nu_2 = \nu$  
and also equal ion-NEMS separation implying in $\kappa_1 = \kappa_2 = \kappa$. 
In this symmetric scenario, the set of eigenvalues (or the spectrum) of  
$\mathsf{ J } \mathcal{H}$ is  
${\rm Spec}_{\mathbb{C}}(\mathsf{ J } \mathcal{H}) = \{\eta_{\pm},\pi_{\pm},\rho_{\pm}\}$, with
\begin{eqnarray}
\eta_{\pm} &=&  \pm i{\omega}_{+}\nonumber\\
\pi_{\pm} &=& \frac{\pm i}{\sqrt{2}} 
        \sqrt{ \omega^2 + {\omega}_{-}^2 + 
        \sqrt{ \left[ \omega^2 - {\omega}_{-}^2\right]^2 
              + 8\kappa^2\omega{\omega}_{-}} }\nonumber\\
\rho_{\pm} &=& \frac{\pm i}{\sqrt{2}} 
        \sqrt{ \omega^2 + {\omega}_{-}^2 - 
        \sqrt{ \left[ \omega^2 - {\omega}_{-}^2\right]^2 
              + 8\kappa^2\omega{\omega}_{-}} },\nonumber\\                                                                               \label{spec}
\end{eqnarray}
where we defined ${\omega}_{\pm}=\nu \pm \Omega$.

It is the $\kappa$-dependence of the quartet $\{\pi_{\pm},\rho_{\pm}\}$ which will determine the type of dynamics. 
Complex conjugate pairs of eigenvalues in ${\rm Spec}_{\mathbb{C}}(\mathsf{ J } \mathcal{H})$ indicate 
rotations whereas a real pair of eigenvalues with opposite signs indicates hyperbolic movement in the phase space, 
the same as squeezing. 
From (\ref{spec}) it is easy to see that the first two eigenvalues $\eta_{\pm}$ are always purely imaginary. 
Besides, if $ \kappa \le \sqrt{\omega(\nu - \Omega)}$, ${\rm Spec}_{\mathbb{C}}(\mathsf{ J } \mathcal{H})$ 
will consist only of pure imaginary eigenvalues indicating rotation in phase space. On the other hand, 
if $ \kappa > \sqrt{\omega(\nu-\Omega)}$, we obtain a ``mixed'' dynamics of rotations and ``squeezing" 
which is then capable of generating 
some entanglement in the system. The same dynamical behavior can be obtained with more compact equations if 
we establish $\omega = \nu -\Omega $. 
Since $\nu >> \Omega$, in general,  this corresponds to the limit of approximate resonance of natural 
frequencies of the ion vibrational motion and NEMS. 
In this case the above spectrum simplifies to  
\begin{equation}
{\rm Spec}_{\mathbb{C}}(\mathsf{ J } \mathcal{H}) = 
\left\{ \pm i\omega_+, 
        \pm i \sqrt{ \omega(\omega +\kappa)}, \pm i \sqrt{ \omega(\omega - \kappa)  }\right\},    \label{spec2}
\end{equation}%
such that by taking $\kappa > \omega$, one will get a mixed dynamics (rotations+squeezing). 
We recall that generating squeezing is a necessary condition to have entanglement from a separable initial 
coherent or thermal state following Gaussian evolutions. Therefore, we shall now fix $\kappa > \omega$. 
The matrix elements of $\mathsf E_t$ necessary to obtain (\ref{ev}) 
for this spectrum can be found in appendix \ref{A1}.%
%
\subsection{Closed system with $T=0$}\label{t0}
In this subsection, we will be considering an initial pure state consisting of the product of 
coherent states for the ions and the NEMS, this leads to the initial CM ${\bm \gamma}_0 = \frac{1}{2} {\bf 1}_6$, 
where ${\bf 1}_6$ is the $6\times 6$ identity matrix. 
We then evolve this initial covariance matrix according to (\ref{ev}). 
In order to facilitate the interpretation of the matrix elements, 
we arrange it in sectors representing each system %
and it is given by
\begin{equation} \label{covtotal}
{\bm \gamma}(t) = \left(
\begin{array}{c|c}
  {\bm \gamma}_{\rm N} & \mathbf{C} \,\,\,\,  - \mathbf{C} \\ \hline
  \mathbf{C}^\top & \raisebox{-5pt}{{\mbox{{${\bm \gamma}_{\rm I}$}}}} \\[0 ex]
 -\mathbf{C}^\top & \\[0 ex]
  \end{array}
\right)      ,                                                                            
\end{equation}
where
\begin{equation}    \label{covions}
{\bm \gamma}_{\rm I} = \frac{1}{2}
\left(\begin{array}{cc} 
\mathbf A_{\rm I} & \mathbf C_{\rm I} \\
\mathbf C_{\rm I}^\top & \mathbf A_{\rm I}
\end{array}\right).      
\end{equation}
The submatrices are shown in appendix \ref{A1} . 
It is worthwhile to notice 
the independence of the system CM (\ref{covtotal})
on $\omega_{+}$ which is the only parameter in (\ref{spec2}) that contains information 
on the ion-ion coupling constant $\Omega$. This independence may be understood as a 
consequence of the fact that $\omega_{\pm}$ appears solely in the imaginary eigenvalues (rotations) 
not affecting the second moments. We are then in a regime where \textit{all entanglement is due to 
the coupling of the ions to the NEMS and not due the coupling between the ions}.

Looking at the CM of the Ion-Ion subsystem, eq.(\ref{covions}),  
and it's elements in appendix \ref{A1},  one can see that 
\begin{equation}\label{sp1}
\mathbf{A}_{\rm I} = {\mathsf I}_2 - \mathbf{C}_{\rm I}, 
\end{equation}
which has profound consequences 
in this system: it is then possible to detect and quantify 
entanglement doing only local measurements, 
since the correlations described by $\mathbf{C}_{\rm I}$ 
are given in terms of the CM of the local modes, {\it i.e.}, 
$\mathbf{A}_{\rm I}$. 

\begin{figure}[htb!]
 \centering\includegraphics[width=1.0\columnwidth]{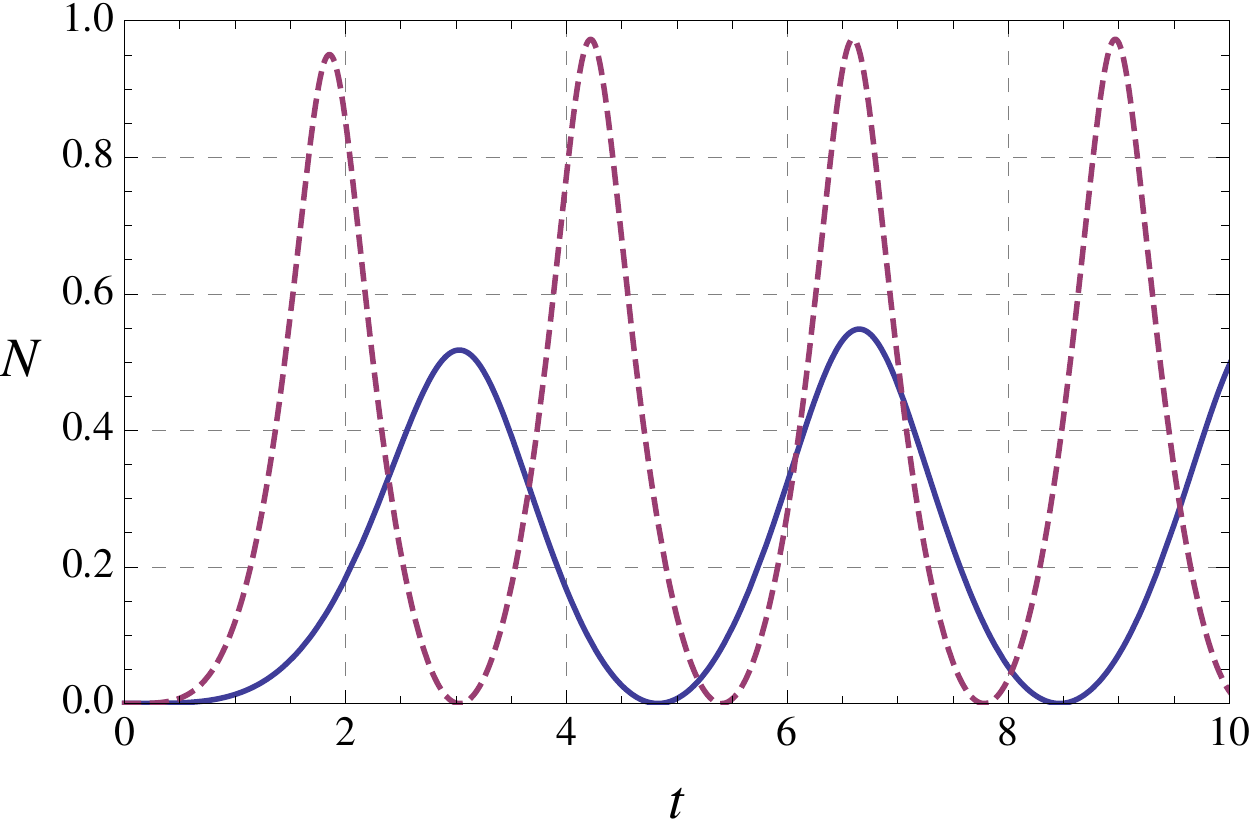}
 \caption{Logarithmic negativity for the ions when the system is initially prepared in a 
 product of coherent states. Two coupling constants between the ions and the NEMS were considered: 
 $\kappa=1$ (solid) and $\kappa=3$ (dashed). Also, $\omega = 0.5$ has been considered.}  
 \label{fig2}
\end{figure}

In figure \ref{fig2}, the NEMS-induced entanglement between the ions is presented as a function of time. 
This plot shows that the stronger the NEMS-ion coupling $\kappa$, the stronger the entanglement between the ions. 
This coupling constant is externally controlled by the experimentalist and $\omega$ is fixed by the NEMS fabrication.
Basically, according to (\ref{kappa}), $\kappa$ can be enhanced by mere increasing of the gate voltage on the NEMS. 
It should be stressed that there is a lower bound on $\kappa$ for this creation of entanglement to happen 
($\kappa$ must be stronger than $\omega$). 
Consequently, figure \ref{fig2} illustrates a central result in this paper. 
Two trapped ions initially uncorrelated and prepared in coherent states can become entangled by 
interacting with a nanoelectromechanical resonator (also prepared in a coherent state) as soon as 
the ion-NEMS coupling achieve a certain value, and this can be controlled by external voltage gate on the NEMS device.

%
\subsection{Closed system with $T\neq0$}
 A thermal state is a thermal equilibrium state of a quantum 
oscillator with frequency $\omega$ in contact with a thermal reservoir at temperature $T$ 
(canonical ensemble). Its density matrix becomes diagonal in the Fock state basis
$| m \rangle$: 
\begin{equation}          \label{term}
\hat \rho_{\!_T} =   \frac{1}{\bar n + 1}
                 \sum_{m = 0}^{\infty} \left( \frac{\bar n}{\bar n + 1} \right)^{\! m } \! 
                 | m \rangle \! \langle m |, 
\end{equation}                                            
where the mean phonon number is
\begin{equation}
\bar n = \left[ {\rm exp}\!\left(-\frac{ \omega }
                                       { {k}_{\!\!_B} T }\right) - 1 \right]^{-1} \ge 0               
\end{equation}
and $ {k}_{\!\!_B} $ is the Boltzman constant. %
The CM of such state is written as 
$ {\bm \gamma}_{\!_T} :=  ( \bar n + \tfrac{1}{2})  {\bf 1}_2$. By defining $ \alpha = 2 \bar n + 1$ 
and still considering the ions in a product of coherent states, the initial CM of the total system is  
$ {\bm \gamma}_0^\alpha =  \tfrac{1}{2} {\rm Diag}(\alpha {\bf 1}_2,{\bf 1}_2,{\bf 1}_2)$ and
its evolution, following (\ref{ev}) and (\ref{spec2}), will be 
\begin{equation}                                                                         \label{covtotalterm} 
{\bm \gamma}_\alpha (t) =  
\left(
\begin{array}{c|c}
  {\bm \gamma}_{\alpha\rm N} & \mathbf{C}_\alpha \,\,\,\,  - \mathbf{C}_\alpha \\ \hline
  \mathbf{C}_\alpha^\top & \raisebox{-5pt}{{\mbox{{${\bm \gamma}_{\alpha\rm I}$}}}} \\[0 ex]
 -\mathbf{C}_\alpha^\top & \\[0 ex]
  \end{array}
\right)
\end{equation}                                                                                  
 with
\begin{equation} \label{covni2}
{\bm \gamma}_{\alpha \rm I} = \frac{1}{2}
\left(\begin{array}{cc} 
\mathbf A^\alpha_{\rm I} & \mathbf C^\alpha_{\rm I} \\
{\mathbf C^\alpha_{\rm I}}^\top & \mathbf A^\alpha_{\rm I}
\end{array}\right).                                                                      
\end{equation}

It is then clear that (\ref{covtotalterm}) and (\ref{covtotal}) have the same structure, but the matrix elements in (\ref{covtotalterm}) are much more involved than in (\ref{covtotal}).  We could check that the former has the nice symmetry of admitting a decomposition as $ \mathbf{G}(t,k,\omega) + \alpha \,\mathbf{H}(t,k,\omega)$, but the
matrices $\mathbf{G}$ and $\mathbf{H}$ are composed by lengthy 
and cumbersome expressions in $t$, $k$ and $\omega$ which does not provide direct physical insight about the problem. For this reason, we will not present them explicitly. 
In the limit of $T\rightarrow 0$ ($\alpha\rightarrow 1$), 
we have checked analytically that ${\bm \gamma}_\alpha (t)\rightarrow {\bm \gamma}(t)$ 
which is given by (\ref{covtotal}).

Just like before in the $T=0$ case described by (\ref{sp1}), it is still possible to detect and quantify locally 
the entanglement of this system because the matrix elements of 
(\ref{covni2}) can be written as  
\begin{equation} \label{sp2}
\mathbf A^\alpha_{\rm I} = \mathsf I_2 - \mathbf C^\alpha_{\rm I}.                                    
\end{equation}

\begin{figure}[htb!]
 \centering\includegraphics[width=1.0\columnwidth]{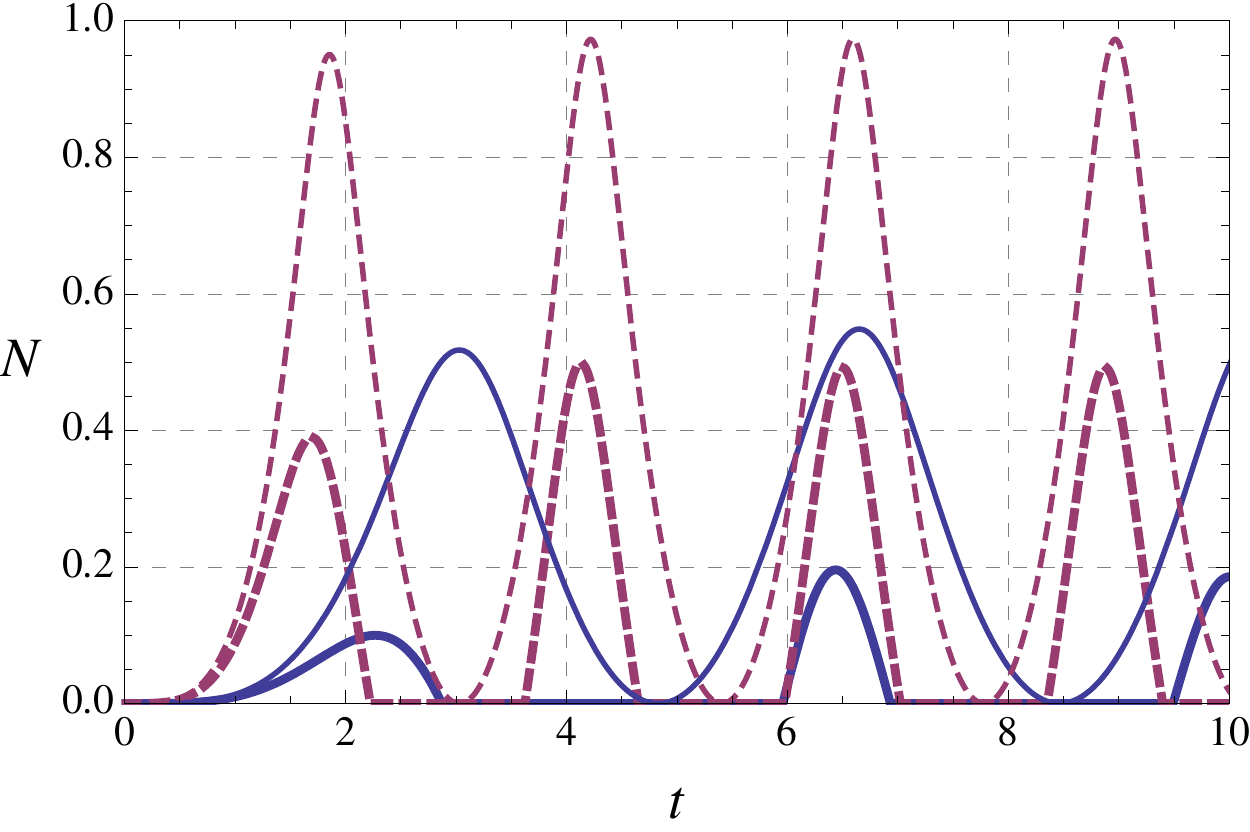}
 \caption{Logarithmic negativity for the ions when the NEMS is prepared in a thermal 
          state characterized by $\alpha$ and the ions are initially prepared in a 
          product of coherent states. 
          Thin lines correspond to zero temperature ($\alpha = 1$) and thick lines correspond 
          to a finite temperature ($\alpha=5$). For the NEMS-ions coupling we used $\kappa=1$ (solid) 
          and $\kappa=3$ (dashed). Also, $\omega=0.5$ has been considered.}
 \label{fig3}
\end{figure}
\begin{figure}[htb!]
 \centering\includegraphics[width=1.0\columnwidth]{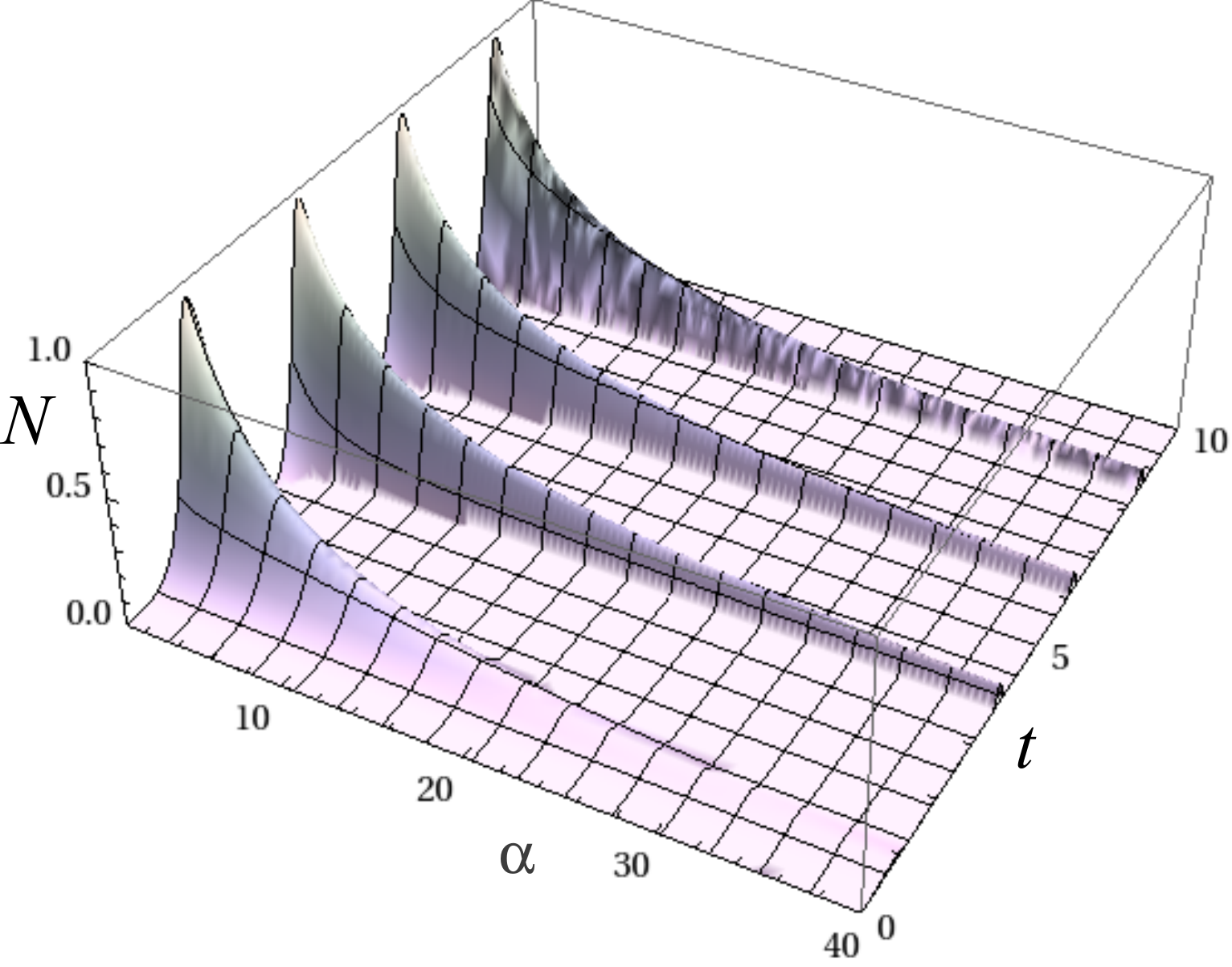}
 \caption{Logarithmic negativity for the ions as a function of time $t$ and $\alpha$ which 
 is an increasing function of the temperature of the NEMS (see main text). The ions were initially 
 prepared in a product of coherent states. The other parameters are $\omega=0.5$ and $\kappa = 3$.} 
 \label{fig4}
\end{figure}

 
Let us now investigate the entangling power of a thermal NEMS. Firstly, let us study the ion-ion 
dynamics for two different temperatures as shown in figure \ref{fig3}. We can see that by 
increasing the temperature (increasing $\alpha$), two interesting trends appear. 
On one side, we can clearly see that the entanglement generated between the ions 
for a finite temperature is smaller than in the $T=0$ case. In a certain way, 
this is expected since thermal fluctuations in the NEMS tend to turn it into a classical object. 
On the other hand, we can see that for a finite temperature there are collapses and revivals 
of entanglement in the dynamics. 
These behaviors can be seen in Fig.\ref{fig4}. 
In general, the NEMS' entangling power degrades asymptotically as $\alpha$ grows, {\it i.e.,} 
as thermal fluctuations in the NEMS becomes dominant over quantum fluctuations. More details can be seen in appendix \ref{A2}.

%
%
\subsection{Open System Dynamics and Tripartite Entanglement Resilience}
%
We now study the case presented in Section \ref{IntEnv} where we briefly 
reviewed how dissipation in the NEMS can be taken into account in this Gaussian system. 
Our main goal is to investigate the resilience of bi- and tripartite entanglements in the system. 
To treat the genuine tripartite entanglement in our system, we use the methods explained in Section \ref{lte}. 
In all the following plots, solid lines refer to the bipartite ion-ion entanglement, 
dashed lines refer to the bipartite NEMS-ion entanglement and dotted lines refer to the 
tripartite entanglement. Also, the initial system preparation is a product of coherent 
states for each system, just like in Section \ref{t0}. We would also like to emphasize 
that the bipartite entanglement between each ion and the NEMS are identical in the frequency 
regimes considered here, as previously explained.

\begin{figure}[htb!]
 \centering\includegraphics[width=1.0\columnwidth]{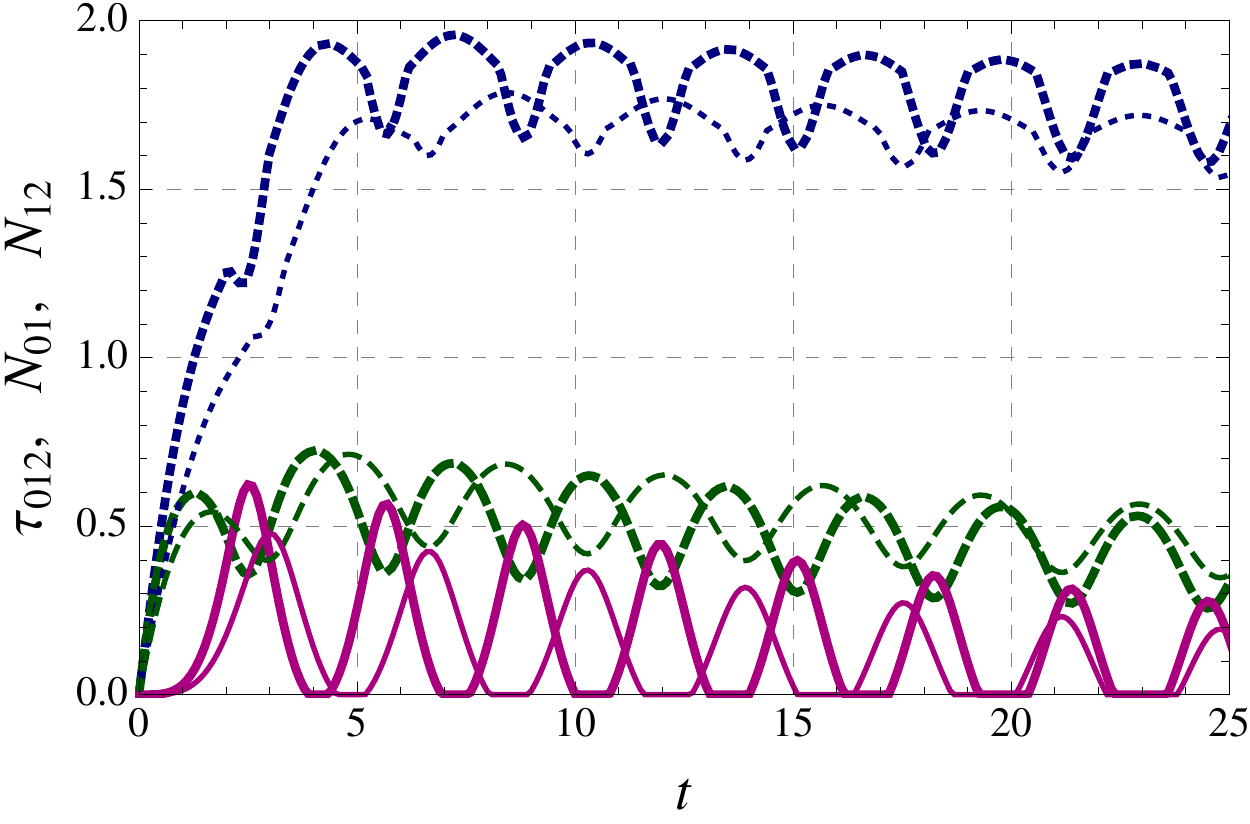}
 \caption{Open system dynamics for bi- and tripartite entanglements in the system 
 consisting of one NEMS coupled to two trapped ions. Thin lines refer to $\kappa=1$ and thick lines to $\kappa=1.5$. 
 The other system parameters are $\zeta=0.01$ and $\omega=0.5$. 
  The mean thermal occupation number 
  in the reservoir is $\bar{N}=4.5$. 
 Dotted lines corresponds to $\tau_{012}$, dashed lines to $N_{01}$ and continuous lines to $N_{12}$.}
 \label{fig5}
\end{figure}

In figure \ref{fig5} we present the open system time evolution of bi- and 
tripartite entanglements for two different NEMS-ion couplings $\kappa$. 
The general behavior does not change with $\kappa$, except the maximal value of the entanglements 
which clearly increases with $\kappa$. From this plot, it is quite clear that the tripartite entanglement 
is much more resilient than the bipartite ones to losses in the NEMS. The latter goes to zero much later 
than the former as we can check numerically.

\begin{figure}[htb!]
 \centering\includegraphics[width=1.0\columnwidth]{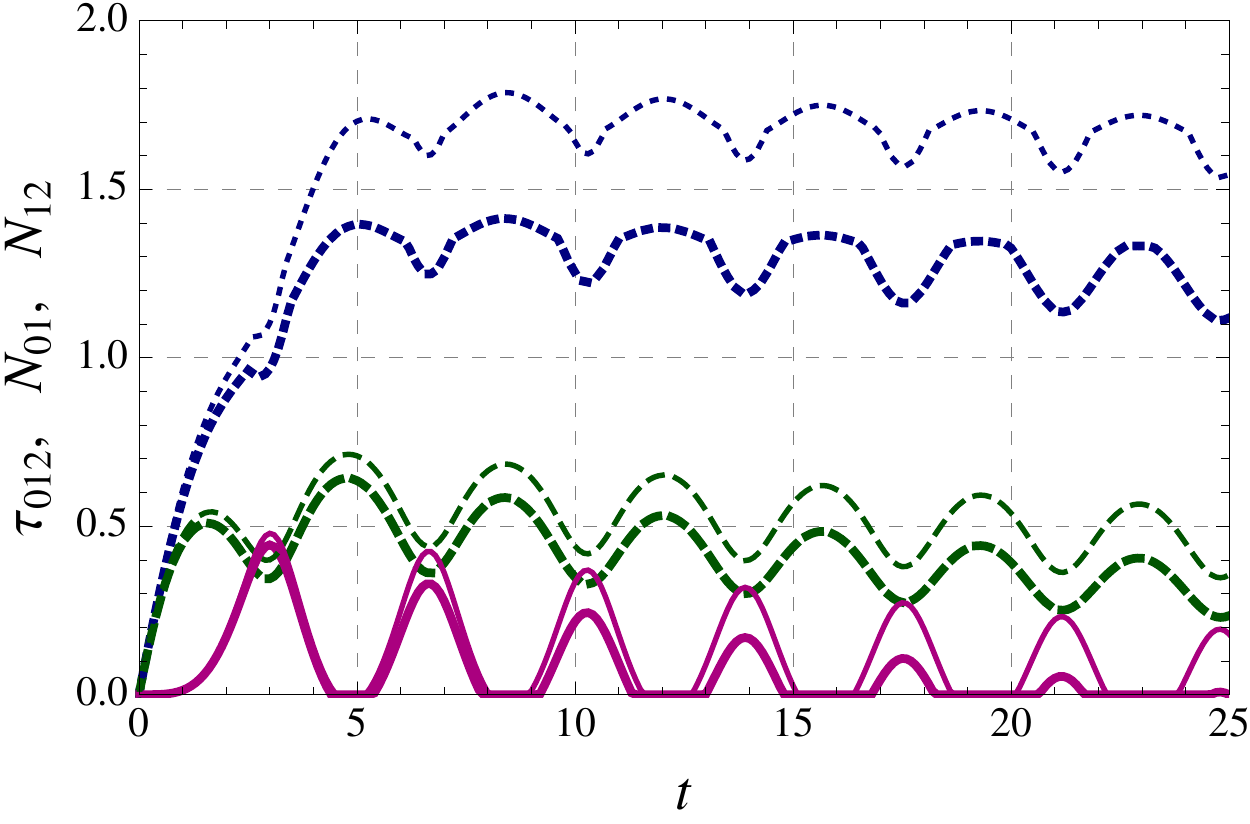}
 \caption{Open system dynamics for bi- and tripartite entanglements in the system 
         for two different NEMS's decay constants. Thin lines refer to $\zeta=0.01$ 
         and thick lines to $\zeta=0.02$. The other system parameters are $\kappa=1$ 
         and $\omega=0.5$.
         The mean thermal occupation number in the reservoir is $\bar{N}=4.5$.
         {\it Idem} fig.\ref{fig5}.}
 \label{fig6}
\end{figure}

In order to further understand this apparent resilience of the tripartite entanglement 
when compared to the bipartite ones, we now present their time evolution for two different 
NEMS decay rates, keeping the reservoir temperature constant. This is presented in figure \ref{fig6}. 
As could be expected, the stronger the decay rates the stronger the suppression of entanglement. 
However, it is still remarkable that the tripartite entanglement takes a much longer time than the
bipartite ones to go to zero, signalizing again a pronounced robustness to coupling to the 
environment. 

\begin{figure}[htb!]
 \centering\includegraphics[width=1.0\columnwidth]{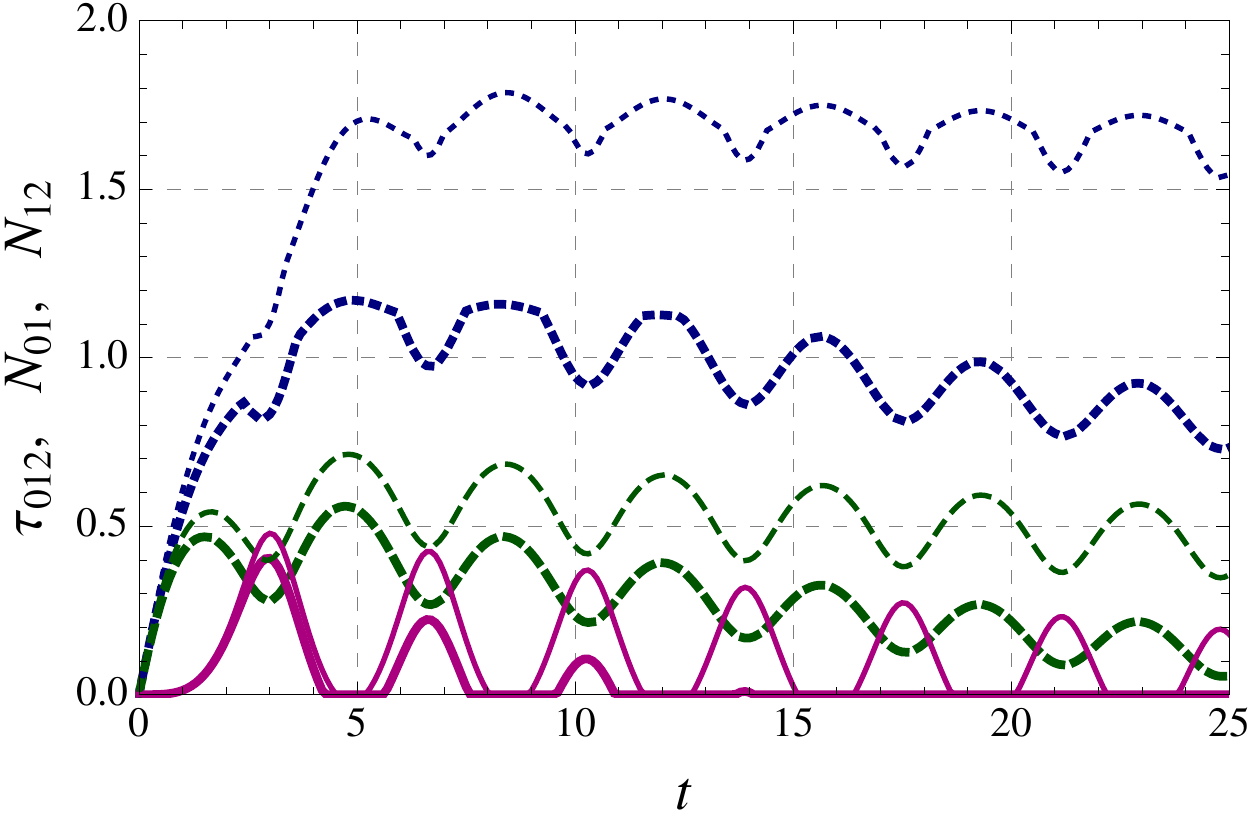}
 \caption{Open system dynamics for bi- and tripartite entanglements in the system 
 for two different temperatures. Thin lines refer to $\bar{N}=4.5$ and thick lines to $\bar{N}=15$. 
 The other system parameters are $ k = 1$, $\omega=0.5$ and $\zeta=0.01$. {\it Idem} fig.\ref{fig5}.}
 \label{fig7}
\end{figure}

Now we analyze the thermal effect on the open system dynamics. The results for two 
different temperatures is shown in figure \ref{fig7}. Although thermal noise severely 
degrades entanglement, as expected, it is still noticeable that tripartite entanglement 
is more robust than the bipartite ones. An increase in temperature is more destructive to 
the bipartite entanglement in our system than to the tripartite one in the sense that the 
latter goes to zero much sooner than the former. 
This robustness is studied in appendix \ref{A2}.
%
\subsection{Conclusions}\label{Conclusion}
We have presented some results concerning entanglement generation in a tripartite 
system consisting of two atomic ions and a nanoscale mechanical resonator. 
We showed that the latter is capable of inducing entanglement in the former 
and we studied the features of such generation under ideal and nonideal conditions. 
In the ideal case we show analytically the precise regime where 
the system becomes entangled or not. The amount of entanglement is controlled 
by the gate voltage applied on NEMS.  
The different trends followed by the bipartite in tripartite entanglements in the system are show in appendix 
\ref{A2} where we present simple expressions fitting the data with good precision. 
For instance, we find that the amount of entanglement created between the ions through their interaction with 
the NEMS follows basically a logarithmic growth with the ion-NEMS coupling constant.
We have also found that the tripartite entanglement is much more resilient to losses 
in the NEMS than the bipartite ones as the latter goes to zero much later than the former. 
We believe our studies may be useful for deepen our understanding about the interaction 
of atomic and nanoscale systems, especially in what concerns the appearing and 
destruction of nonlocal quantum correlations under ideal and nonideal conditions.

\begin{acknowledgments}
During the revision of this work, Professor Kyoko Furuya passed away. Her serenity, kindness and
altruism will stay in our thoughts as a role model for the rest of our lives.
We wish to thank O. P. de S\'a Neto for his valuable 
participation at the beginning of this work and in the formulation of the problem. 
We would also like to thank Mauro Paternostro for his hints on multipartite entanglement analyzes. %
F.N. wishes to thank financial support from FAPESP (Procs. 2009/16369-8). 
K.F. and F.L.S. acknowledge participation as members of the Brazilian National Institute 
of Science and Technology of Quantum Information (INCT/IQ). 
K.F. acknowledges partial support from CNPq  under grant $305568/2010-8$.  
F.L.S. also acknowledges partial support from CNPq under grant $308948/2011-4$. F.N. and F.L.S. thank the Program Science
Without Borders by CNPq Procs. 401265/2012-9 and Grant
No. 150480/2013-0.
\end{acknowledgments}
\appendix
\section{Some analytical matrix elements}\label{A1}
In this appendix, we present the explicit analytical form of some matrix 
elements needed to obtain results in the main body of this manuscript. 

The matrix elements of ${\mathsf E_t}$ (\ref{E}) relative to the spectrum (\ref{spec2}) are given by
\begin{eqnarray}
{\mathsf E_t}_{11} & = & {\mathsf E_t}_{44} = \tfrac{1}{2} ( \cosh\tilde{\omega}t + \cos\bar\omega t ),    \nonumber \\
{\mathsf E_t}_{12} & = & {\mathsf E_t}_{21} =  {\mathsf E_t}_{45} = {\mathsf E_t}_{54} 
            =  \tfrac{\sqrt{2}}{4}( \cosh\tilde{\omega}t - \cos\bar\omega t ),           \nonumber \\
{\mathsf E_t}_{13} & = & {\mathsf E_t}_{31} =  {\mathsf E_t}_{46} = {\mathsf E_t}_{64}  = - {\mathsf E_t}_{12},                       \nonumber \\ 
{\mathsf E_t}_{22} & = & {\mathsf E_t}_{55} = \tfrac{1}{4} ( \cosh\tilde{\omega}t + \cos\bar\omega t 
                                                                + 2\cos\omega_{+}t  ) ,  \nonumber \\
{\mathsf E_t}_{23} & = & {\mathsf E_t}_{32} = {\mathsf E_t}_{56} = {\mathsf E_t}_{65}  \nonumber \\
                    &=& \tfrac{1}{4} (2 \cos\omega_{+} t - \cos\bar\omega t 
                                               - \cosh\tilde{\omega}t )   ,              \nonumber \\
{\mathsf E_t}_{33} & = & {\mathsf E_t}_{66}  
            = \tfrac{1}{4} ( 2 \cos\omega_{+} t + \cos\bar\omega t 
                                                + \cosh\tilde{\omega}t )  ,              \nonumber \\
{\mathsf E_t}_{14} & = & \tfrac{1}{2} (  \tfrac{\omega}{\bar\omega}   \sin\bar\omega t
                              + \tfrac{\omega}{\tilde\omega} \sinh\tilde\omega t ),      \nonumber \\
{\mathsf E_t}_{15} & = & {\mathsf E_t}_{24} 
            = \tfrac{\sqrt{2}}{4}(-\tfrac{\omega}{\bar\omega} \sin\bar\omega t 
                                 + \tfrac{\omega}{\tilde\omega} \sinh\tilde\omega t ) ,  \nonumber \\
{\mathsf E_t}_{16} & = & {\mathsf E_t}_{34} = -{\mathsf E_t}_{15},                       \nonumber \\
{\mathsf E_t}_{25} & = & {\mathsf E_t}_{36}   
            = \tfrac{1}{4} (    \tfrac{\omega}{\bar\omega} \sin\bar\omega t 
                            +   \tfrac{\omega}{\tilde\omega} \sinh\tilde\omega t 
                            + 2 \sin \omega_{+}t) ,                                      \nonumber \\  
{\mathsf E_t}_{26} & = & {\mathsf E_t}_{35} 
            = - \tfrac{1}{4}(\tfrac{\omega}{\bar\omega} \sin\bar\omega t 
              + \tfrac{\omega}{\tilde\omega} \sinh\tilde\omega t 
              - 2 \sin \omega_{+}t),                                                     \nonumber \\
{\mathsf E_t}_{41} & = & \tfrac{1}{2} (- \tfrac{\bar\omega}{\omega} \sin\bar\omega t 
                              + \tfrac{\tilde\omega}{\omega} \sinh\tilde\omega t ),      \nonumber \\
{\mathsf E_t}_{42} & = & {\mathsf E_t}_{51}   
            = \tfrac{\sqrt{2}}{4} ( \tfrac{\bar\omega}{\omega} \sin\bar\omega t 
                                  + \tfrac{\tilde\omega}{\omega} \sinh\tilde\omega t ),  \nonumber \\
{\mathsf E_t}_{43} & = & {\mathsf E_t}_{61} = - {\mathsf E_t}_{42},                                                 \nonumber \\ 
{\mathsf E_t}_{52} & = & {\mathsf E_t}_{63} 
            = \tfrac{1}{4}( \tfrac{\tilde\omega}{\omega} \sinh\tilde\omega t 
                          - \tfrac{\bar\omega}{\omega} \sin\bar\omega t 
                          - 2 \sin \omega_{+}t),                                         \nonumber\\ 
{\mathsf E_t}_{53} & = & {\mathsf E_t}_{62}  
            =  \tfrac{1}{4}(\tfrac{\bar\omega}{\omega} \sin\bar\omega t 
             - \tfrac{\tilde\omega}{\omega} \sinh\tilde\omega t - 2 \sin \omega_{+}t),  \nonumber 
\end{eqnarray}
where $\tilde{\omega} \equiv \sqrt{\omega(\kappa-\omega)}$, $\bar{\omega} \equiv \sqrt{\omega(\kappa+\omega)}$ 
and $\omega_{+} := \Omega+\nu$.

The submatrices necessary to write the covariance matrix (\ref{covtotal}) are given by
\begin{eqnarray*}
{\bm \gamma}_N &=& \frac{1}{2}\left(
\begin{array}{cc} 
1 + a(t)
  + b(t)    & 
  c(t) + d(t) \\
c(t) + d(t) & 
1 + a'(t)
  + b'(t)
\end{array}\right);                                                                      \nonumber \\
\mathbf A_{\rm I} &=& \left( 
\begin{array}{cc} 
1 + \frac{1}{2}[a(t)+ b(t)]        & 
    \frac{1}{2}[c(t)+ d(t)] \\
\frac{1}{2}[c(t)+ d(t)] &
1 + \frac{1}{2}[a'(t)+ b'(t)]\end{array}\right) ;                              \nonumber \\
\mathbf C_{\rm I} &=& - \frac{1}{2} \left( 
\begin{array}{cc} 
a(t) + b(t) & c(t)+ d(t) \\
c(t)+ d(t)  & a'(t)+ b'(t)
\end{array}\right) ;                                                                     \nonumber \\                                                               
\mathbf{ C } &=& -\frac{\sqrt{2}}{4}\left( 
\begin{array}{cc} 
a(t)-b(t)   &  c(t)-d(t) \\
c(t)-d(t)   & a'(t)-b'(t) 
\end{array}\right),                                                                      \nonumber
\end{eqnarray*}
where
\begin{eqnarray}
a(t)&=&\frac{\omega^2 - \bar\omega^2  }{2 \bar \omega^2   } \sin^2 \bar{ \omega } t, \\
a'(t) &=&\frac{\bar\omega^2 - \omega^2  }{2  \omega^2} \sin^2 \bar{ \omega } t ,\\
b(t) &=&\frac{\omega^2 + \tilde\omega^2}{2 \tilde{\omega}^2} \sinh^2 \tilde{\omega } t,\\
b'(t) &=& \frac{\tilde\omega^2 + \omega^2}{2  \omega^2} \sinh^2 \tilde{\omega } t,\\
c(t) &=& \frac{\omega^2 - \bar{\omega }^2  }{4\omega\bar{  \omega }} \sin  2  \bar{\omega } t ,\\
d(t) &=& \frac{\omega^2 + \tilde{\omega }^2}{4\omega\tilde{\omega }} \sinh 2 \tilde{\omega } t.
\end{eqnarray}
\section{Study of General Trends}\label{A2}
We use this appendix to further analyze results presented in the main text. Analytical expressions for the symplectic eigenvalues used to evaluate entanglement are quite lengthy and it is extremely difficult to extract the general trends seen in the plots from limits and approximations of these expressions. Recognizing the importance of trying to understand the observed behavior through simple yet approximate expressions, we resort to numerics. We employed numerical routines to find the best fitting for a giving fitting function picked by us. In order to favor physical intuition in the open system Markovian dynamics induced by dissipation in the NEMS, our approach was in this case to enforce fitting with exponential functions, even though they might not be the most precise choice. By doing this, we were able, for instance, to assess the robustness of tripartite or bipartite entanglements facing dissipation in the NEMS.

We start by deepen the analyzes of the unitary dynamics of entanglement presented in Fig.\ref{fig2}. 
Looking  at the CM of the Ion-Ion subsystem, eq.(\ref{covions}), 
and the relation (\ref{sp1}), it is then possible to show that the smallest symplectic 
eigenvalue of (\ref{covions}) can be written as 
\begin{equation}\label{se1}
\mu_1^{\!\top_{\!\!B} } = \sqrt{ 1 - 2 \Lambda^{+}_{\mathbf{C}_{\rm I}} }
                        = \sqrt{ 2 \Lambda^{-}_{\mathbf{A}_{\rm I}} -1  },
\end{equation}%
where $\Lambda^{+}_{\bf G}$ and $\Lambda^{-}_{\bf G}$ 
denotes the  biggest and the smallest ordinary (non symplectic) eigenvalues 
of an arbitrary matrix ${\bf G}$. The condition for separability                          
\footnote{For a two mode system, following \cite{cre},   
is possible to show that there can be only one symplectic 
eigenvalue smaller than one, and the necessary and sufficient 
condition for separability is described as 
$\mu^{\!\top_{\!\!B}}  \ge 1$. }                    
is expressed in this situation as 
$\Lambda^{+}_{\mathbf{C}_{\rm I}} \ge 0$ and 
is possible to obtain the period of oscillation 
(a function of $\kappa$ and $\omega$) as the distance 
between two consecutive zeros of $\det \mathbf{C}_{\rm I}$. 
The local maximum of entanglement
will again be a function of $\kappa$ and $\omega$, 
and it can be found  by calculating the maximum eigenvalue 
$\Lambda^{+}_{\mathbf{C}_{\rm I}}$.  
These eigenvalues are given in terms of transcendental functions, 
and the solutions can only be determined numerically. 

\begin{figure}[htb!]
\centering\includegraphics[width=1.0\columnwidth]{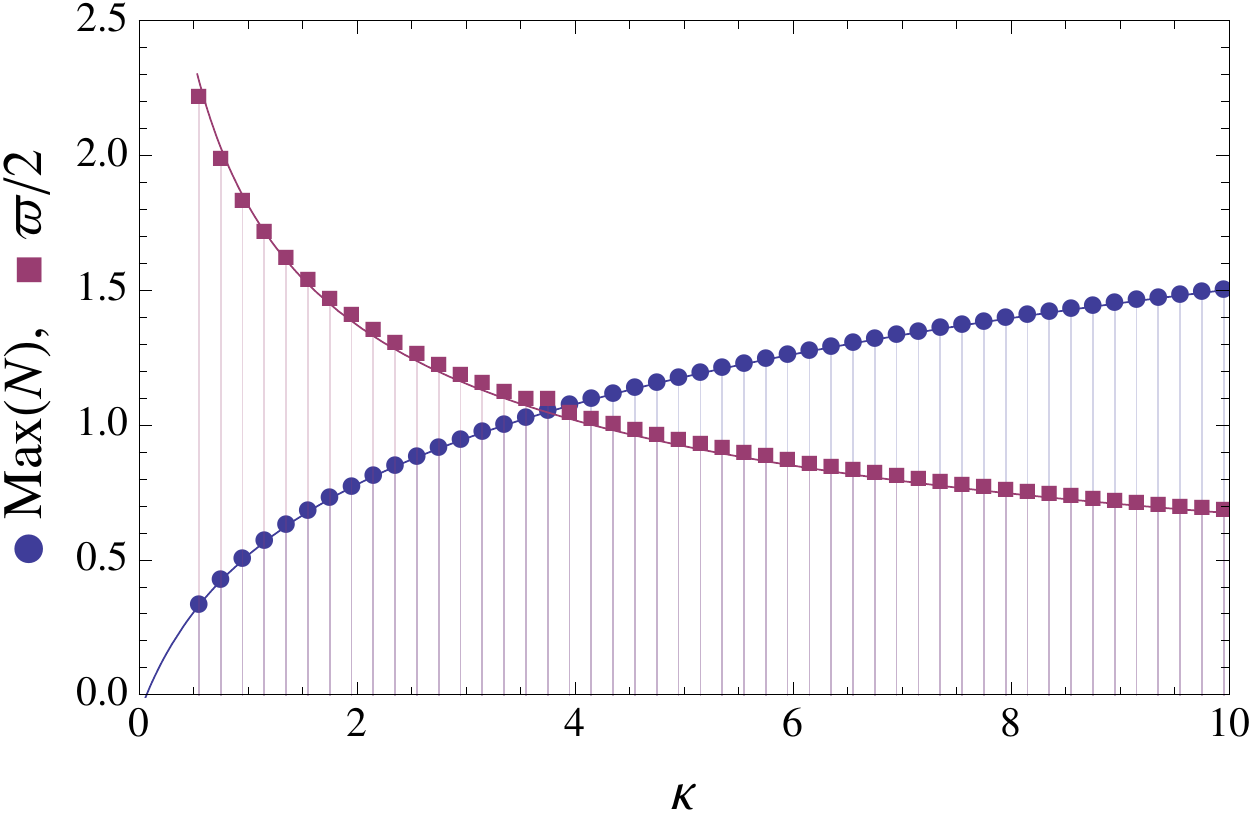}
\caption{First local maximum of the logarithmic negativity (circles) and the 
period of the oscillation (squares) as a function of $\kappa$ for 
the same situation of fig.\ref{fig2}. The continuous line is the numerical fit 
performed with (\ref{fit1}), (\ref{fit1t}), and parameters (\ref{fit1p}). 
The relative error of these fits are smaller than $5\%$. }         \label{figap1}
\end{figure}

In Fig. \ref{fig2}, one can see that by increasing $\kappa$, the entanglement generated between the ions 
also increases while the period of oscillations decreases. 
To a better understanding of these important trends, we fit our results with the functions
\begin{equation}\label{fit1}
{\rm Max}(N) = b \, {\rm ln }(c \, \kappa + d),
\end{equation}
\begin{equation}\label{fit1t}
\varpi =  (e \, \sqrt{\kappa} + f)^{-1} ,
\end{equation}
where ${\rm Max}(N)$ is the first local maximum of the logarithmic negativity and $\varpi$ the 
period of the oscillation. The fit presented in Fig. \ref{figap1} was performed setting
\begin{equation}\label{fit1p}
b = 0.5, \,\,\, c = 2, \,\,\,  
d = 0.9  \,\,\, e = 0.1 \,\,\, {\rm and } \,\,\, f = 0.4 \, . 
\end{equation}
The relative error of these fits are smaller than $5\%$ for the range $\kappa \in [0,40]$,
and $\omega = 0.5$ was picked in accordance with (\ref{fig2}). 
The cases with different values of $\omega$ follow the same behavior as long as one sets $\kappa \rightarrow \kappa/\omega $ and $t\rightarrow \omega t$, and this can be seen directly from matrix elements of (\ref{covions}) given in appendix A.
%

Now we move to the situation 
where the entanglement depends also on the temperature through the 
parameter $\alpha$. Taking (\ref{sp2}) into account, it is again possible to write something similar to (\ref{se1}). Explicitly we get
\begin{equation}
\mu_1^{\!\top_{\!\!B} } \! = \! \sqrt{ 2 \Lambda^{-}_{\mathbf{A}_{\rm I}^\alpha} \! - \! 1  } 
\! = \! \sqrt{\! a_1 \!+\! a_2 \alpha\! - \!\sqrt{\!a_3 \! + \! a_4 \alpha \!+\! a_5 \alpha^2} },  \,\,  
\end{equation}
where $a_i$  are combinations of functions of $t,\omega,\kappa$.  
This gives the dependence of the logarithmic negativity 
(\ref{logsym}) on $\alpha$ and it comes from the evaluation of the eigenvalues of 
$\mathbf{A}_{\rm I}^\alpha$ when considering that any element of this matrix 
can be splitted as $g_{ij} + \alpha \, h_{ij} $ as commented in the main text. 
This is the exact situation shown in fig.\ref{fig3}, for given choices of parameters.

\begin{figure}[htb!]
\centering\includegraphics[width=1.0\columnwidth]{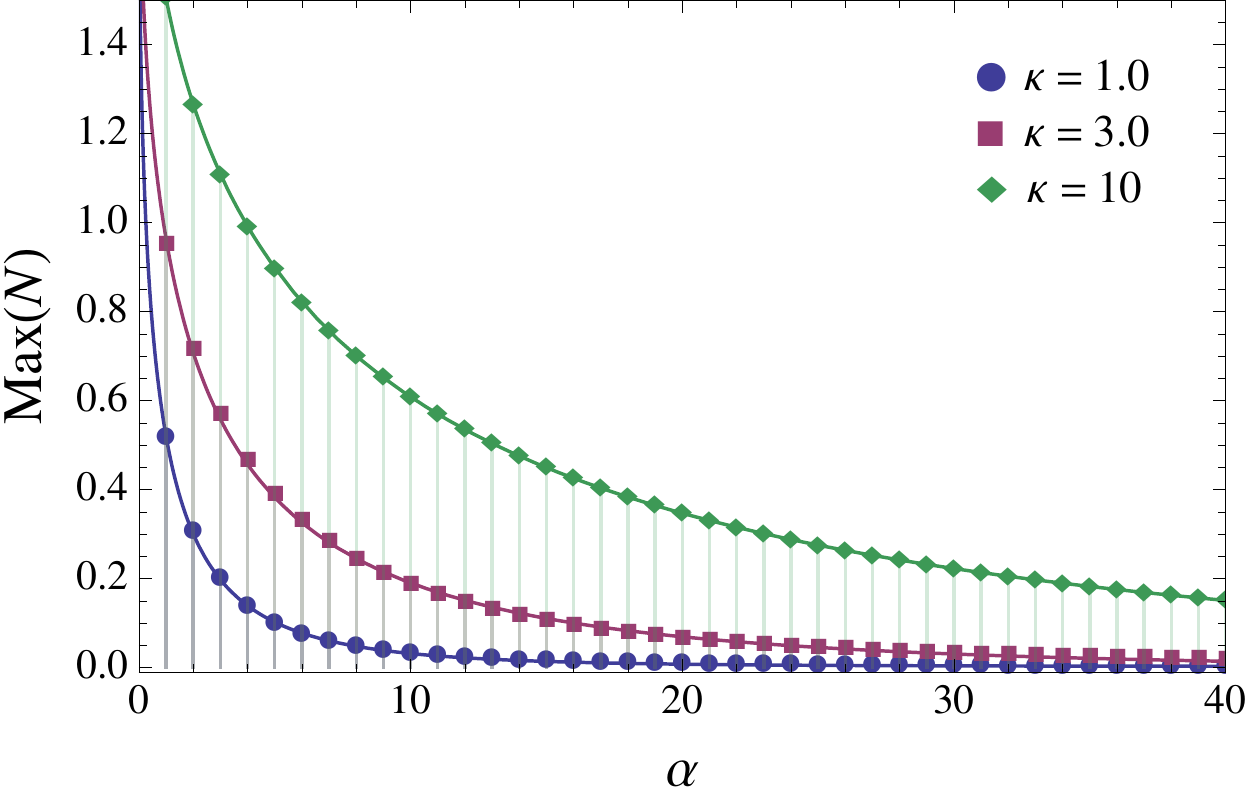}
\caption{First local maximum of the logarithmic negativity as a function of $\alpha$. 
We consider here the same situation of fig.\ref{fig3} with $\kappa = 1$ (circles), 
$\kappa = 3$ (squares) and $\kappa =10$ (diamonds). 
We also plot a numerical fit for both set of points, 
the relative error of these fits are smaller than $5\%$ 
for all points and $\omega = 0.5$.}                       \label{figap2}
\end{figure}

Another interesting question is to follow 
the first maximum of entanglement in fig.\ref{fig3} 
as the temperature changes. The position in time of the first maximum also changes when $\alpha$ is varied, but the period is not changed. This is explained below.
In fig.\ref{figap2}, we present this dependence, 
and one can clearly see that in this case a exponential in $ \sqrt{\alpha}$ approximates 
very well the observed behavior. We fit this curve with
\begin{equation} \label{fit2}
{\rm Max}(N) = b(\kappa) \, {\rm e }^{- c(\kappa) \sqrt{\alpha}} 
\end{equation}
and
\begin{equation}
b(1) = 1.91, \,\,\, c(1) = 1.3, \,\,\,  
b(3) = 2.08  \,\,\, {\rm and } \,\,\, c(3) = 0.75 \, . 
\end{equation}
We believe that these simple formulas (\ref{fit1}) and (\ref{fit2}) may be valuable for the experimentalist to formulate strategies to extract information about physical quantities such as temperature $\alpha$ or ion-NEMS coupling $\kappa$ through measurements of entanglement. Moreover, given the special forms (\ref{sp1}) and (\ref{sp2}), entanglement between the ions can then be extract from local measurements on one ion only, and consequently this applies to (\ref{fit1}) and (\ref{fit2}). Finally, the period of oscillations in fig.\ref{fig3} does not depend on $\alpha$ 
by the simple reason that time appears only in the 
symplectic matrix ${\mathsf E}_t$ given by (\ref{ev}), and dependence on $\alpha$ appears only in ${\bm \gamma}_0$. Therefore, the period is still given by (\ref{fit1t}). For larger values of $\kappa$ or $\alpha$, the behavior of the first maximum of entanglement remains the same, 
a decreasing exponential in $\sqrt{\alpha}$ and a growing function with $\kappa$, 
suggesting that the negativity will always be produced for a given choice of $\kappa$ and $\alpha$, see fig.\ref{figap2}.
%

In the open dynamics we were mainly interested in assessing the robustness or fragility of the bipartite and tripartite entanglements. 
The procedure to calculate entanglement in Gaussian systems relies on the 
determination of symplectic eigenvalues of matrices. 
In spite of the fact that the determination of the covariance matrix is in essence analytic as discussed in the main text,
the expressions for the symplectic eigenvalues turn out to be huge combinations of non factorable
exponentials, trigonometric  and hyperbolic functions. 
The cumbersomeness of these expressions prevents us to explore analytically 
some of their consequences. In part, this is due to the fact that now the symmetry present in (\ref{se1}) 
is broken due to dissipation in the NEMS, and the analysis becomes purely numerical. 
%

\begin{figure}[htb!]
\centering\includegraphics[width=1.0\columnwidth]{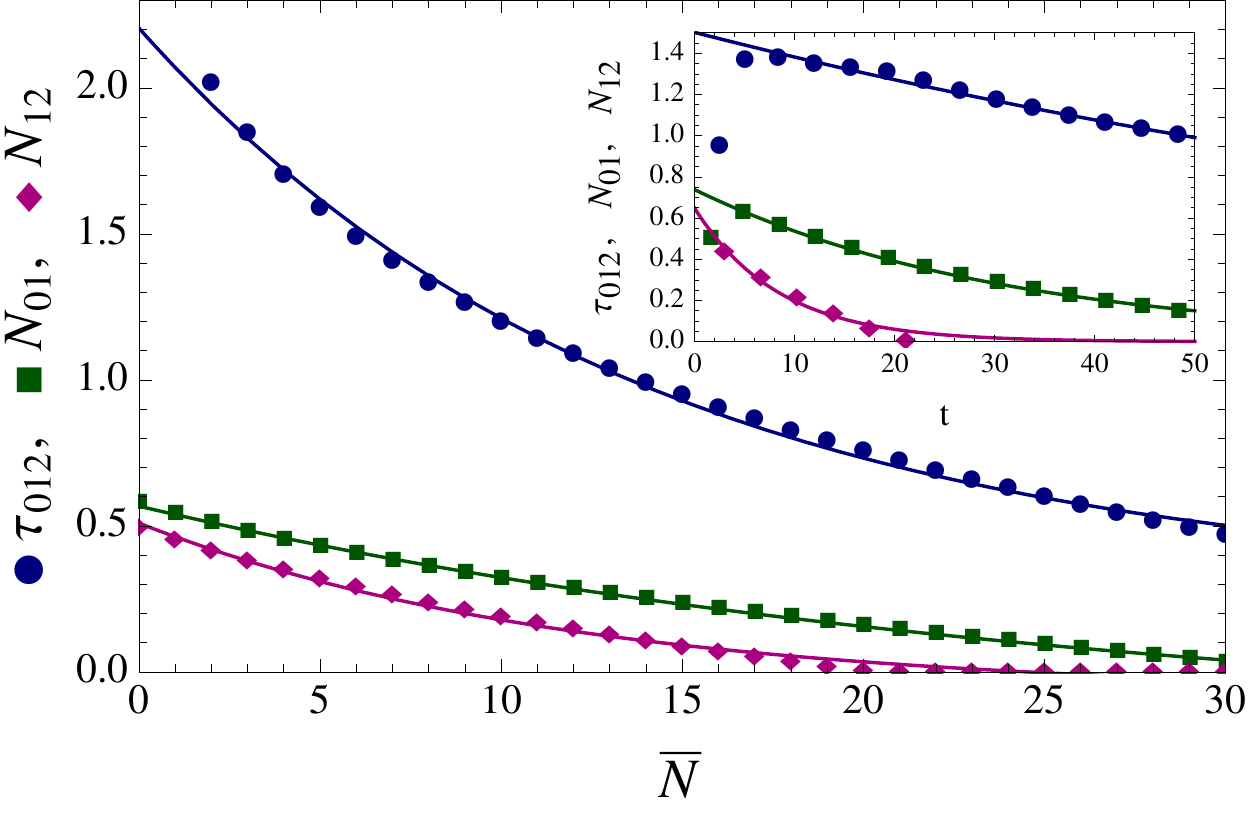}
\caption{Behavior of entanglement as a function of $\bar{N}$ for 
the fixed instant of time $t =10$.  
We consider here $\kappa=1$, $\zeta=0.01$ and $\omega=0.5$. 
In the inset we plot the local maxima of entanglement 
as function of time for $\bar N = 10$.
We show the tripartite entanglement's measure (circles), 
the negativity of NEMS-ion subsystem (squares) 
and the bipartite ion-ion negativity (diamonds).
We also plot a numerical fit for all set of points.}          \label{figap3}
\end{figure} 

First let us analyze how the amount of entanglement in a given time is affected by the mean 
occupation thermal number of the reservoir $\bar N$. 
In fig.\ref{figap3}  an example of this dependence and the fitting with decaying exponentials is presented. The fitting used in this case was
\begin{eqnarray}
\tau_{012} &=& a_1 + b_1 {\rm e}^{-c_1 \, \bar{N}} , \,\,\, a_1 = 0.29 , b_1 = 1.92 , c_1 = 0.07 \nonumber \\ 
N_{01}      &=& a_2 + b_2 {\rm e}^{-c_2 \, \bar{N}} , \,\,\, a_2 = - 0.21 , b_2 = 0.78 , c_2 = 0.04 \nonumber \\   
N_{12}      &=& a_3 + b_3 {\rm e}^{-c_3 \, \bar{N}}, \,\,\,  a_3 = - 0.08 , b_3 = 0.58 , c_3 = 0.08  \nonumber \\
\end{eqnarray} 
From the values of $c_1,c_2$ and $c_3$, it is pretty clear that the temperature of the NEMS reservoir affect both tripartite and bipartite entanglements in essentially the same way. However, in the inset we present the sequence of local maxima for a fixed reservoir temperature (fixed $\bar{N}$), and it seems that the tripartite entanglement is more robust in the sense that it decays slowly than the bipartite counterparts. Interestingly, this is indeed the case as understood from the best exponential fitting 
\begin{eqnarray}
\tau_{012} &=& a_1 {\rm e}^{-b_1 \, t} , \,\,\, a_1 = 1.50 , b_1 = 0.008  \nonumber \\ 
N_{01}      &=& a_2 {\rm e}^{-b_2 \, t} , \,\,\, a_2 = 0.74 , b_2 = 0.032 \nonumber \\   
N_{12}      &=& a_3 {\rm e}^{-b_3 \, t}, \,\,\,  a_3 = 0.65 , b_3 = 0.120 
\end{eqnarray} 
It is interesting to check whether or not the decay time scale for the entanglements are comparable to the ones of the Langevin dynamics which is around $\zeta^{-1}$. In the case of the inset of Fig.\ref{figap3}, we set $\zeta=0.01$, and it is then clear that the decay of entanglement do not necessarily occur at such precise scale but it decays at a time which is of the same order (around $10^1-10^{2}$). 

\begin{figure}[htb!]
\centering\includegraphics[width=1.0\columnwidth]{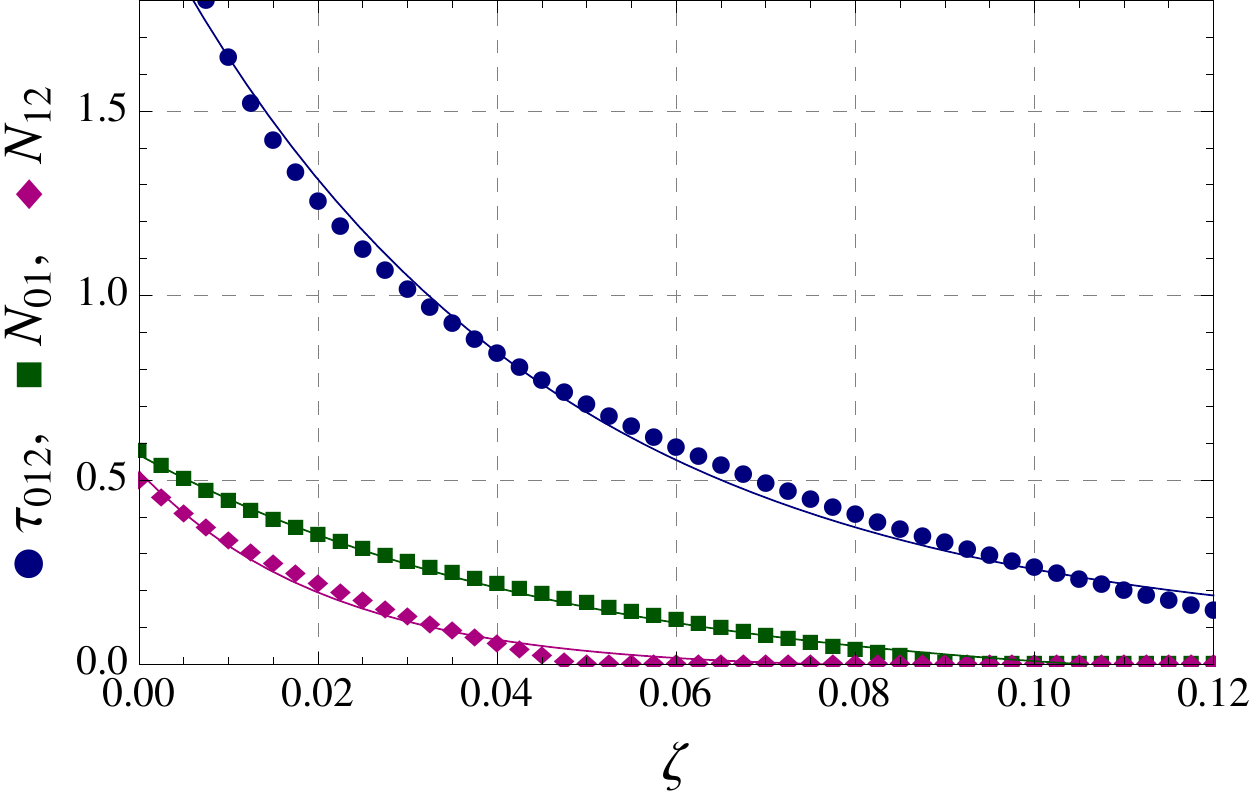}
\caption{ Behavior of entanglements measure as a function of $\zeta$ for 
a fixed instant of time for $t =10$.  
We consider here $\kappa=1$, $\bar{N} = 4.5$ and $\omega=0.5$. 
 {\it Idem} fig.\ref{figap3}.}          \label{figap4}
\end{figure} 
 %

We carry on the throughly analyzes of this open system case, by looking into the dependence of the entanglements with the damping constant constant $\zeta$. This study is presented in Fig.\ref{figap4}, and the fitting used was
\begin{eqnarray}
\tau_{012} &=& a_1 + b_1 {\rm e}^{-c_1 \, \zeta} , \,\,\, a_1 = 0.07 , b_1 = 2.00 , c_1 = 23.60 \nonumber \\ 
N_{01}      &=& a_2 + b_2 {\rm e}^{-c_2 \, \zeta} , \,\,\, a_2 = -0.07 , b_2 = 0.64 , c_2 = 20.75 \nonumber \\   
N_{12}      &=& a_3 + b_3 {\rm e}^{-c_3 \, \zeta}, \,\,\,  a_3 = - 0.01 , b_3 = 0.54 , c_3 = 47.20  \nonumber \\
\end{eqnarray} 
%
\begin{figure}[htb!]
\centering\includegraphics[width=1.0\columnwidth]{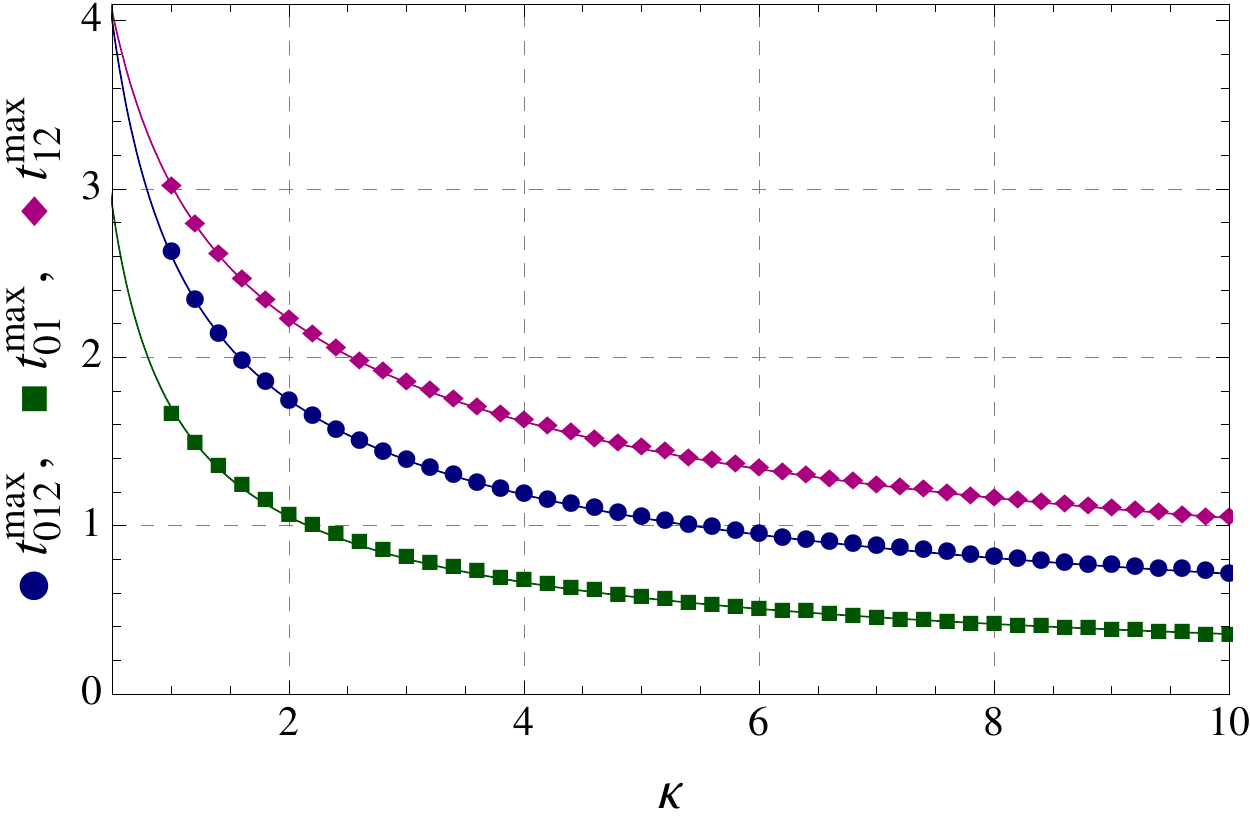}
\caption{Time in which the entanglement measures attain the maximum 
as a function of $\kappa$. 
We consider here $\bar{N} = 4.5$, $\zeta = 0.1$ and $\omega=0.5$. 
The relative error of these fits are smaller than $5\%$. 
{\it Idem} fig.\ref{figap3}.}          \label{figap5}
\end{figure} 

Since all entanglement in the system is created by 
the coupling $\kappa$, it might be interesting to see how fast the maximum entanglement is achieved as a function of $\kappa$. This is shown in Fig.\ref{figap5}. In this case, the exponential is not very appropriate and we fit with
\begin{eqnarray}
t^{{\rm max}}_{012} &=& a_1 + \frac{1}{ b_1 + c_1 \sqrt{k}} ,  \,\,\, 
t^{{\rm max}}_{01}     =    a_2 + \frac{1}{ b_2 + c_2 \sqrt{k}},  \nonumber \\
t^{{\rm max}}_{12}  &=&  a_3 + \frac{1}{ b_3 + c_3 \sqrt{k}} ,
\end{eqnarray}
where $a_1 = -0.04, b_1 = -0.06 , c_1 = 0.44$, 
       $a_2 = -0.10 , b_2 = -0.20,   c_2 = 0.75$,
       $a_3 = -0.02 , b_3 = 0.05  \,\,\, {\rm and} \,\,\, b_3 = 0.28$. 
       
\newpage

\end{document}